\documentclass{natureprintstylev4}

\usepackage{astjnlabbrev-nature}

\usepackage{amssymb}
\usepackage{mathptmx}
\usepackage{amsmath}	
\usepackage{gensymb}
\usepackage{color}
\usepackage{enumitem}

\usepackage{epsfig}
\usepackage{color}
\usepackage{graphicx}
\usepackage{longtable}
\usepackage{hyperref}

\usepackage[labelsep=endash]{caption}

\newcommand{\citep}{\cite}
\newcommand{\citet}{\cite}

\title{Quiescent Ultra-diffuse galaxies in the field originating from backsplash orbits}

\author{Jos\'e A. Benavides$^{1,2}$\thanks{E-mail: jose.benavides@unc.edu.ar},
Laura V. Sales$^{3}$,
Mario. G. Abadi$^{1,2}$,
Annalisa Pillepich$^{4}$,
Dylan Nelson$^{5}$, 
Federico Marinacci$^{6}$,
Michael Cooper$^{7}$,
Ruediger Pakmor$^{8}$, 
Paul Torrey$^{9}$,
Mark Vogelsberger$^{10}$ and
Lars Hernquist$^{11}$
.}

\begin{document}

\maketitle

\begin{affiliations}
\item Instituto de Astronom\'ia Te\'orica y Experimental, CONICET-UNC, Laprida 854, X5000BGR, C\'ordoba, Argentina.
\item Observatorio Astron\'omico de C\'ordoba, Universidad Nacional de C\'odoba, Laprida 854, X5000BGR, C\'ordoba, Argentina.
\item Department of Physics and Astronomy, University of California, Riverside, CA, 92521, USA.
\item Max-Planck-Institut f\"{u}r Astronomie, K\"{o}nigstuhl 17, 69117 Heidelberg, Germany. 
\item Institut f\"ur theoretische Astrophysik, Zentrum f\"ur Astronomie, Universit\"at Heidelberg, D-69120 Heidelberg, Baden-W\"urttemburg, Germany. 
\item Department of Physics and Astronomy "Augusto Righi", University of Bologna, I-40129 Bologna, Italy. 
\item Department of Physics and Astronomy, University of California, Irvine,
4129 Frederick Reines Hall, Irvine, CA 92697, USA. 
\item Max-Planck-Institut f\"ur Astrophysik, Karl-Schwarzschild-Str. 1, D-85748 Garching, Germany. 
\item Department of Astronomy, University of Florida, 211 Bryant Space Science Center, Gainesville, FL 32611, USA. 
\item Department of Physics, Massachusetts Institute of Technology, Cambridge, MA 02139, USA. 
\item Institute for Theory and Computation, Harvard-Smithsonian Center for Astrophysics, Cambridge, MA 02138, USA. 
\end{affiliations}
\\

\textbf{Ultra-diffuse galaxies (UDGs) are the lowest-surface brightness galaxies known, with typical stellar masses of dwarf galaxies but sizes similar to larger galaxies like the Milky Way\cite{vanDokkum2015a}. The reason for their extended sizes is debated, with suggested internal processes like angular momentum \cite{Amorisco2016}, feedback \cite{DiCintio2017, Chan2018} or mergers \cite{Wright2021} versus external mechanisms \cite{Safarzadeh2017,Carleton2019,Jiang2019,Tremmel2020} or a combination of both \cite{Sales2020}. Observationally, we know that UDGs are red and quiescent in groups and clusters \cite{Koda2015, Yagi2016} while their counterparts in the field are blue and star-forming \cite{Greco2018b, Rong2020b,Barbosa2020,Tanoglidis2021}. This dichotomy suggests environmental effects as main culprit. However, this scenario is challenged by recent observations of isolated quiescent UDGs in the field \cite{MartinezDelgado2016,Roman2019,Prole2021}. Here we use $\Lambda$CDM (or $\Lambda$ cold dark matter, where $\Lambda$ is the cosmological constant) cosmological hydrodynamical simulation to show that isolated quenched UDGs are formed as backsplash galaxies that were once satellites of another galactic, group or cluster halo but are today a few Mpc away from them. These interactions, albeit brief, remove the gas and tidally strip the outskirts of the dark matter haloes of the now quenched seemingly-isolated UDGs, which are born as star-forming field UDGs occupying dwarf-mass dark matter haloes. Quiescent UDGs may therefore be found in non-negligible numbers in filaments and voids, bearing the mark of past interactions as stripped outer haloes devoid of dark matter and gas compared to dwarfs with similar stellar content.}\\

\begin{figure*}[!ht]
\centerline{
\includegraphics[width=0.9\textwidth]{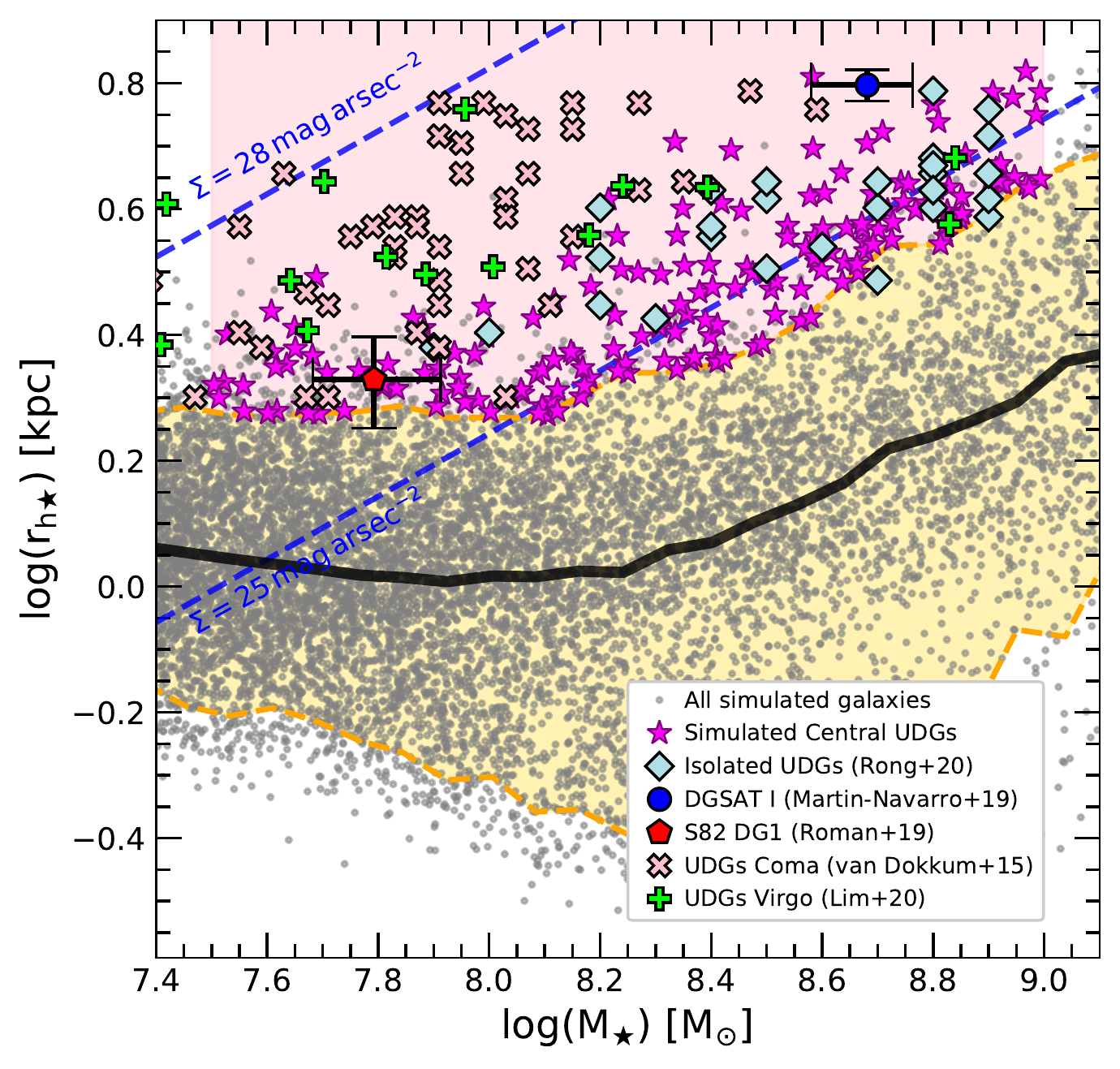}
}
\caption{{\bf Definition of the UDG sample}. Stellar mass versus size ($\rm{M_{\star}}$ vs $\rm{r_{h \star}}$) relation for all simulated galaxies in the mass range $\rm{ \log (M_{\star} / M_{\odot}) =  [7.4, 9.1] } $ in the TNG50 simulation (grey dots). Thin blue dashed curves indicate lines of constant surface brightness assuming a mass-to-light ratio equal to 1. The solid black line indicates the median size at fixed $M_\star$ for the simulated galaxies. Yellow dashed curves show the 5th and 95th percentiles, with the shaded yellow region in between highlighting the sample of normal galaxies. Our sample of field UDGs (magenta stars) is defined as central galaxies with $\rm{ \log (M_{\star} / M_{\odot}) =  [7.5, 9] } $ and stellar size above the 95th percentile (pink shaded region). Several observational data are shown in black edged symbols, where we transform 2D sizes $R_{\rm eff}$ to 3D assuming $\rm{r_{h \star} = 4/3 R_{\rm eff}}$. Light blue diamonds indicate star-forming UDGs in low-density environments \cite{Rong2020b}; the dark blue circle is the relatively isolated DGSAT I [\cite{MartinNavarro2019}]; the red pentagon is UDG S82-DG-1, an isolated quiescent UDG \cite{Roman2019}. For comparison, we also show UDGs in the Virgo cluster \cite{Lim2020} (green crosses) and the Coma cluster \cite{vanDokkum2015a} (pink x-symbols). Our UDG definition agrees well with observational samples, in particular for those in low density environments. }
    \label{fig:sample}
\end{figure*}

Ultra-diffuse galaxies (UDGs) in groups and clusters are characterized by a puzzling wide range of dark matter and globular cluster content \cite{vanDokkum2018,Lim2018,Doppel2021}, thick disk-like shapes \cite{Koda2015,Mancera-Pina2019}, old stellar populations \cite{Ferre2018} and no substantial gas component. Their quiescence is not surprising given the high density environments they populate. On the other hand, for the few quenched UDGs that have been discovered in the field, the mechanism responsible for removing the gas and halting star-formation remains unknown. On the theory side, progress requires high resolution cosmological simulations that are able to resolve the myriad of environments and physics involved in this problem; from the formation of isolated dwarfs in their haloes, to their interactions with filaments, groups and clusters. Such simulations have only recently become possible, with the TNG50 simulation --  used here -- among those with the highest resolution available \cite{Pillepich2019,Nelson2019TNG50}.  \\


\begin{figure*}[!ht]
\centerline{
\includegraphics[width=0.9\textwidth]{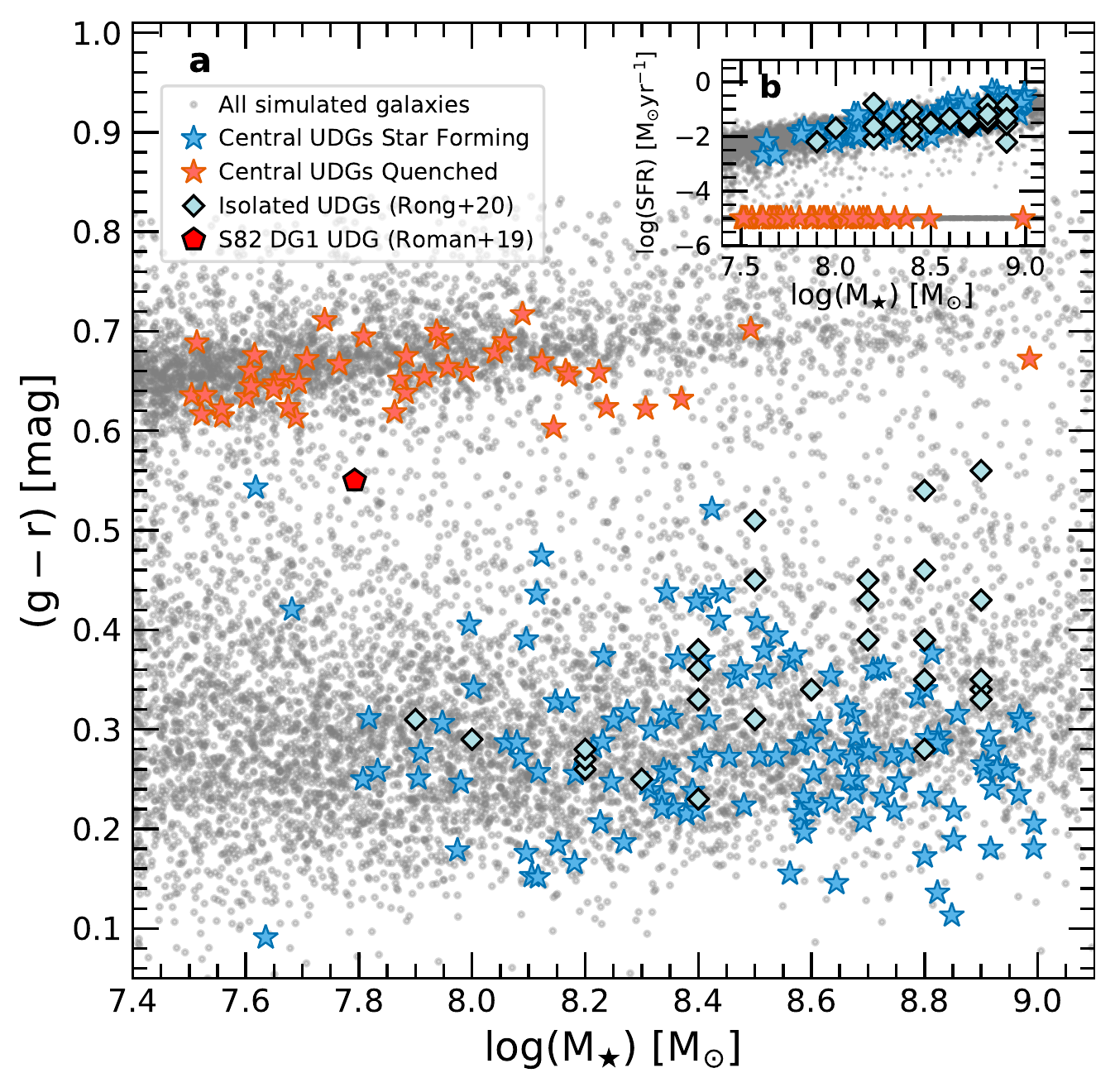}
}
\caption{{\bf Dichotomy in colour and star formation rate of field UDGs}. Panel \textbf{a}, colour ($g$ - $r$) as a function of stellar mass for all simulated galaxies in this mass range (grey dots) and field UDGs (starred symbols). Most field UDGs are blue, but about a quarter of the sample populates the ``red sequence". These colours correlate with star formation rates (SFRs, small inset, panel \textbf{b}), where blue UDGs are star-forming and red UDGs are quiescent (SFR is zero in these objects but has been artificially shifted to $10^{-5} \, \rm M_{\odot} \rm yr^{-1} $ for plotting purposes). Coloured symbols correspond to available observational data: isolated star-forming UDGs [\cite{Rong2020b}] with light blue diamonds and S82-DG-1 [\cite{Roman2019}] with red pentagon.}
    \label{fig:color}
\end{figure*}

We use the stellar mass–size relation defined by all simulated galaxies (Fig.~\ref{fig:sample}, grey dots) to define our sample of field UDGs as those central galaxies (excluding satellites) with stellar mass in the dwarf range ($\rm log(\rm M_{\star}/\rm M_\odot) = [7.5,9]$, shaded pink region) and a stellar size above the $95$th percentile at a given mass (magenta stars). Our definition overlaps with observational samples of UDGs\cite{Rong2020b, MartinNavarro2019, Roman2019,Lim2020}. We study the origin of the extended sizes of these galaxies elsewhere (J.A.B. et al., manuscript in preparation) but a brief summary is given in Fig.~\ref{fig:spin} in the Supplementary Information. \\

Galaxy colours ($g$-$r$) of our simulated UDGs in Fig.~\ref{fig:color} show a clear bimodality: most central UDGs are in the `blue cloud', suggesting young stellar populations as expected for dwarfs in the field, while $23.7\% $ of our simulated UDGs are along the red sequence. The mass distribution is non-uniform, with red UDGs being more common towards the lower masses, where also, at the same mass, field UDGs have a higher fraction of red objects than normal dwarfs in the field (See upper panel of Fig.~\ref{fig:spin} in the Supplementary Information). The inset panel shows that their colours correlate with their star formation rates, with the blue UDGs occupying the ``main sequence" of star-forming galaxies and the red UDGs showing negligible star formation today. \\

\begin{figure*}
\centering
\includegraphics[width=1\textwidth]{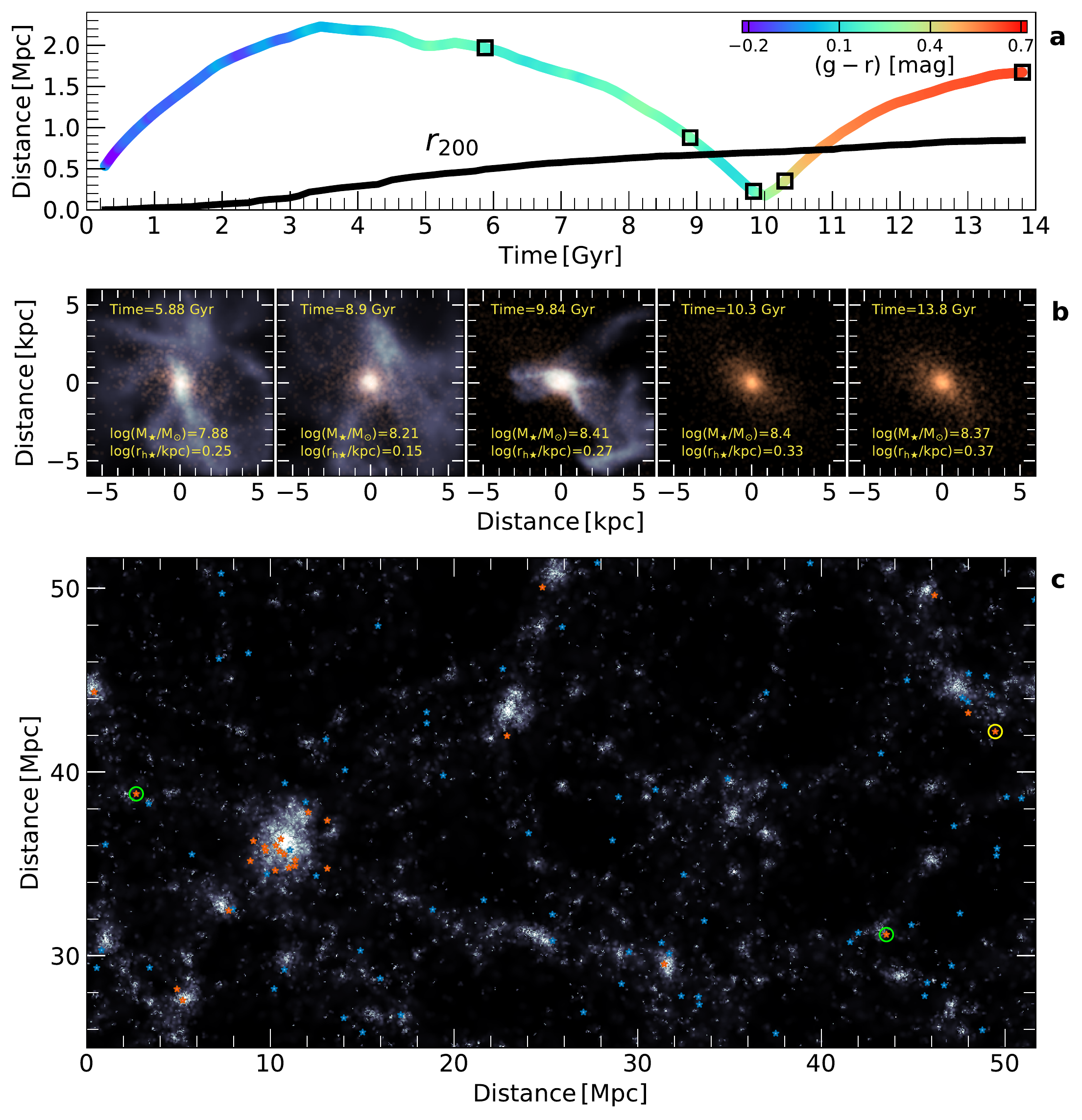}
\caption{{\bf Formation of red UDGs in backsplash orbits and their location in the Universe}. \textbf{a}, Example orbit of one of our quiescent UDG around its temporary (group-size) host halo, whose virial radius is indicated by the black line. The orbit is colour coded according to the instantaneous ($g$-$r$) colour in the dwarf (see colour bar in the top right) and shows that reddening starts right after pericenter. \textbf{b}, snapshot view of the stellar (red) and gas (blue) content of the UDG in different epochs along its orbit (times highlighted by the black squares in the upper panel). The gas is fully removed while approaching the pericenter, resulting in posterior aging and reddening of the stellar population. \textbf{c}, location of blue and red UDGs (starred symbols) in part of the simulation box. Structure is shown in the gray map as traced by all galaxies in the halo catalog. Red UDGs are spatially more clustered than blue ones, but some might exist even in very low density regions. For instance,  the open green circles correspond to UDGs that are backsplash objects of galactic haloes with $\rm{\log(M_{200}/M_{\odot}) \thickapprox 12.5}$. The single yellow circle indicates the most extreme case of an ejected UDG located at $\sim 3.35$ Mpc from its host.}
\label{fig:orbit_box}
\end{figure*}

A close inspection of the histories of our red quiescent UDGs reveals a factor in common: they have all been satellites of another system in the past but are today central galaxies in the field. The top panel of Fig.~\ref{fig:orbit_box} shows an example of the orbit of one of our red UDGs. This dwarf  interacted $\sim 4$ billion years ago with a group that has a virial mass $\rm{M_{200}(z=0) \thickapprox 6.46 \times 10^{13} \; \rm M_\odot}$ but is found today $\sim 1.5$ Mpc away, more than twice the virial radius of the group (virial quantities refer to the radius enclosing $\rm{200}$ times the critical density of the Universe). The colour coding of the orbit, reflective of the dwarf colour at each time, shows that the reddening starts already as it falls into the group and accelerates after the pericentric passage. The images of the simulated UDG (middle row) clearly show that its gas is removed as it approaches the pericenter, explaining its quiescence and redness today in the field. The stellar size is not largely affected by the interaction, our red UDGs were all already extended before infall (see Fig.~\ref{fig:evol_features} in the Supplementary Information).\\ 

Objects in such external orbits, which are found far beyond the virial radius of their hosts, are known as backsplash galaxies \cite{Balogh2000}, and are a natural consequence of the hierarchical assembly in $\rm{\Lambda CDM}$. Our red UDGs are backsplash objects of systems in a wide range of virial masses, including galaxy-sized haloes with $\rm{M_{200} \thickapprox 2 \times 10^{12}\; \rm M_\odot}$ to galaxy clusters, and are today on average at $\rm{2.1 r_{200}}$ from those systems, or $1.7 \pm 0.7$ Mpc, but can reach as far as $3.35$ Mpc in some cases (see Fig.~\ref{fig:distances} in the Supplementary Information). In the large majority of cases, ($64.3\%$), the system responsible for the quenching and the launching beyond the virial radius is the same, with the remaining cases being ``pre-processing'', meaning that the UDG was first quenched in a moderate mass host which subsequently fell into a more massive system responsible for the energetic orbit. \\

A section of the simulated box with the location of red and blue UDGs is illustrated in Fig.~\ref{fig:orbit_box} (panel c), highlighting the red UDGs that are backsplash objects of galaxy-size haloes (green circles), located mainly in low-density regions of the Universe. Red field UDGs cluster more than the blue ones, but they are all at substantial distances from their once-hosts. On average, the interactions occurred $\sim$ 5.5 Gyr ago and were moderately quick, with red UDGs spending typically 1.5  Gyr (median) within the virial radius of the systems they are backsplash of (see Fig.~\ref{fig:hist_times} in the Supplementary Information).\\

Interestingly, there are extreme cases where the close pericenter passage of the UDG results in its total ejection from the system in a way reminiscent of those in multiple-body interactions \cite{Sales2007b}. Our most extreme UDG resides $\sim 3.35$ Mpc away from its host and would appear as an extremely isolated object in a void-like environment (see yellow circle in bottom panel of Fig.~\ref{fig:orbit_box}). This UDG fell in as part of a galaxy-size group into a group-size halo with $\rm{M_{200}(z=0) = 3.36 \times 10^{13} \; \rm M_\odot}$ and was ejected more than $6$ Gyr ago after its first pericenter (see Fig.~\ref{fig:ejected} in the Supplementary Information).\\

\begin{figure*}
\centering
\includegraphics[width=0.96\textwidth]{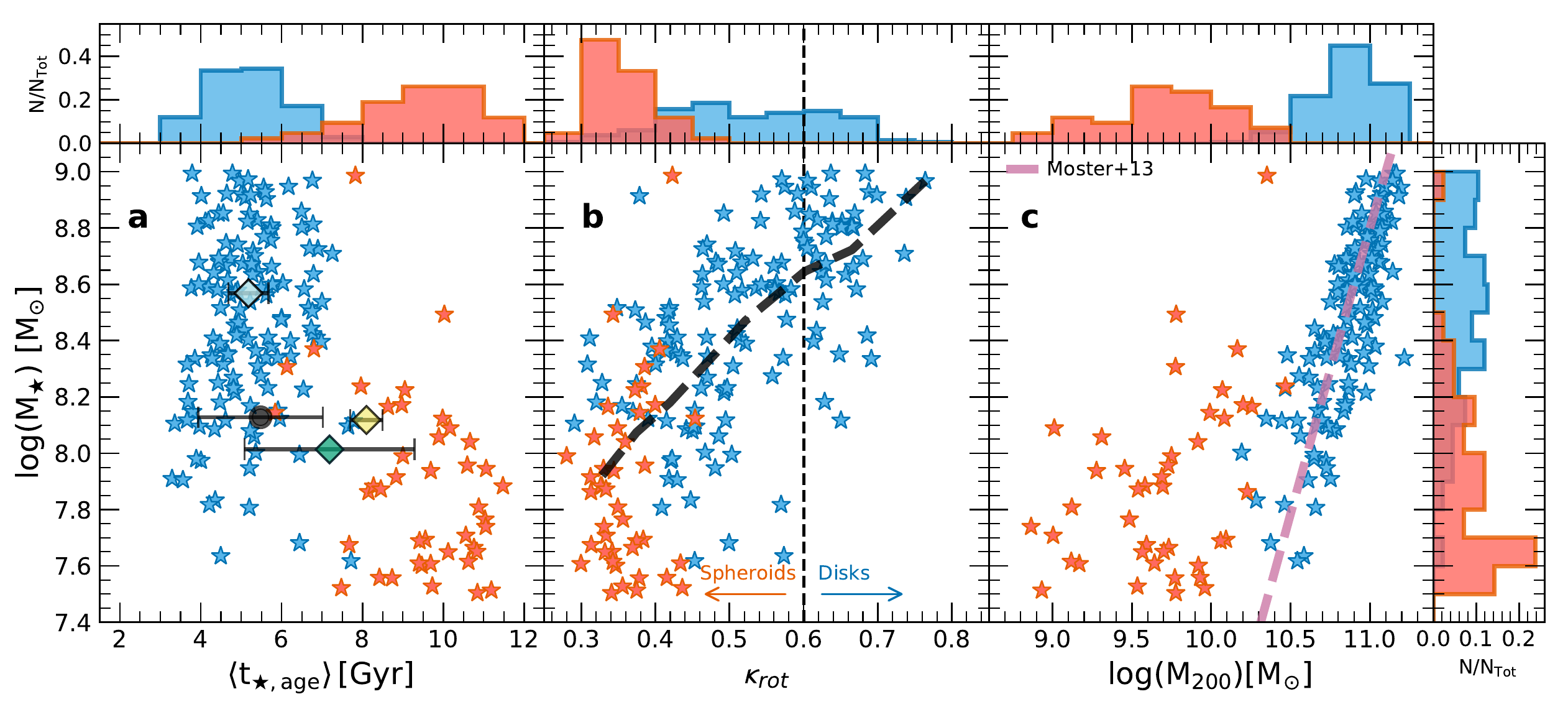}
\caption{{\bf Predicted properties of field red vs. blue UDGs}. Panel \textbf{a}, average age of the population (mass-weighted) as a function of stellar mass. Red UDGs are older than the blue UDGs (average for red and blue are  $\rm{5.21 \pm 0.99 \, Gyr}$ and $\rm{9.40 \pm 1.38 \, Gyr}$, respectively). Blue UDGs have ages consistent with the overall sample of simulated normal (non-UDG) field dwarfs in the same mass range, as shown by the gray symbol with errorbars corresponding to 5th - 95th percentiles. Available observational data is shown for comparison (star-forming UDGs: [\cite{Rong2020b}] with light blue diamond; DGSAT I: [\cite{MartinNavarro2019}] with yellow diamond and an average of UDGs in the Coma cluster \cite{Ferre2018} with the green diamond). Panel \textbf{b}, morphology quantified by the parameter $\kappa_{\rm rot}$, where $\kappa_{\rm rot} > 0.6$ (indicated by the vertical thin dashed line) highlights disk-dominated objects (see Methods). Red UDGs have typically more spheroidal morphology (lower $\kappa_{\rm rot}$ values) than the blue population at similar mass. Black thick dashed curve shows the median $\kappa_{\rm rot}$ at a given stellar mass. Panel \textbf{c}, virial mass ($\rm{ M_{200}}$) - stellar mass relation. Red UDGs have stripped outer haloes and show today smaller virial masses at fixed stellar mass. The pink dashed curve corresponds to the abundance-matching model by [\cite{Moster2013}]. Histograms, showing normalized numbers ($\rm{N/N_{tot}}$) of UDGs, along all axes are included to ease the comparisons.}
\label{fig:features}
\end{figure*}

This scenario for the formation of quiescent UDGs in the field has a number of observational implications. First, the stellar populations are old due to the quenching during the backsplash interaction. As shown in the left panel of Fig.~\ref{fig:features}, blue UDGs are comparatively younger, characterized by an extended star formation history as argued in the case of observed field UDGs \cite{Barbosa2020}, and consistent with the overall simulated population of field dwarfs (gray symbol). Note that the ages inferred for isolated UDGs in observations are mostly in agreement with our blue UDG population.\\ 

Second, the morphologies of red UDGs are always more spheroid-dominated than their blue counterparts with similar stellar mass, which might show spheroid or disc structure (see middle panel of Fig.~\ref{fig:features}), in agreement with previous work \cite{Cardona-Barrero2020}. Here, morphology is quantified by the $\kappa_{\rm rot}$ parameter (Methods). We predict a shift towards early-type morphologies for red UDGs (low $\kappa_{\rm rot}$) which is consistent with the picture where satellite galaxies are preferentially spheroid-dominated due to transformations induced by the environment \cite{Joshi2020}. \\

Third and most important, backsplash galaxies have been stripped to some degree of their mass during the tidal interaction with their past host system. While blue UDGs form in dwarf-mass haloes with virial mass in the range $\rm log(M_{200}/M_\odot) = [10.3-11.2]$, red UDGs at the same stellar mass show smaller virial masses, with a median $\rm log(M_{200}/M_\odot) = 9.73$ (right panel in Fig.~\ref{fig:features}) due to this interaction. Red UDGs in the field should be clear outliers compared to predictions from abundance-matching models \cite{Moster2013}. The stripping occurs mostly in the outer halo, where the dark matter density profile of red UDGs falls more steeply than the unperturbed blue UDG population (see Fig.~\ref{fig:rho_dm} in the Supplementary Information). Unfortunately, the inner stellar velocity dispersion of red and blue UDGs -- a possible observable -- is statistically indistinguishable in our simulation.\\ 

A fourth implication in this scenario is that red UDGs are fully devoid of halo gas, which was all removed via ram pressure \cite{GunnGott1972, Abadi1999} along with the inner gas during the interaction with their hosts. We have checked that no gas is re-accreted in these dwarfs, in contrast with the gas mass $M_{\rm gas} = 10^8$ to $10^{10}\; \rm M_\odot$ predicted in the haloes of blue UDGs in the field (this includes gas with distance (in kpc) of $\rm{2 r_{h \star} < r/\rm kpc < r_{200}}$, where $\rm{r_{h \star}}$ is the stellar half-mass radius). A promising way to study the circumgalactic medium of these galaxies down to very low column densities is to use background quasars to provide different absorption lines-of-sight across a halo \cite{Tumlinson2017}. Although this would be prohibitive on an individual UDG basis, a statistical detection (or lack of thereof) might be achieved once a sufficiently large number of red field UDGs is found. Neutral gas and $H_\alpha$ studies of red UDGs should also confirm a lack of gas in their interstellar medium.\\

There are a few observational detections of quiescent UDGs in low density environments and they seem consistent with the picture emerging from our analysis. One of the first reported non-cluster UDGs is DGSAT I [\cite{MartinezDelgado2016}], which is located in the filament of the Pisces-Perseus supercluster. This is in excellent agreement with our predictions, where most red field UDGs are nearby but outside groups and clusters. DGSAT I lacks gas (as measured by $H_\alpha$ [ref.\cite{MartinezDelgado2016}]) and has a relatively old stellar population ($8.1$ $\pm$ $0.4$ Gyr mass-weighted age, \cite{MartinNavarro2019}) which is also within the range of properties predicted by our simulations.\\ 

Another interesting object is S82-DG-1, an extremely isolated quenched UDG in a nearby void \cite{Roman2019}. Its isolation has been used to favour internal effects such as feedback to explain the possible origin of UDGs, rather than due to high-density environmental effects. Here, we argue that S82-DG-1 fits the characteristic expected for our simulated population of passive UDGs that were satellites of a galactic-size host. S82-DG-1 is located at $\sim 55$ kpc in projection and at a redshift-distance less than $\rm{ \Delta v = 145}$ km $\rm s^{-1}$ from NGC 1211, a lenticular galaxy with stellar mass $\rm{M_{\star} \thickapprox 1 \times 10^{10} \; \rm M_\odot}$ (see Methods). Three of our simulated red UDGs have been backsplash objects in galaxy-mass haloes $\rm{M_{200} < 10^{13}\; M_\odot}$ and are found today $\sim 650$ kpc from their hosts. Moreover, $12$ red isolated UDGs ($28.5\%$) were quenched in galactic environments ($\rm{M_{200} < 10^{13}\; M_\odot}$). Although the exact distance of S82-DG-1 to NGC 1211 is unknown, our analysis provides support for the possible external nature of quenching in S82-DG-1 induced by NGC 1211. The old stellar population inferred for S82-DG-1, $6$ Gyr of age \cite{Roman2019}, is in excellent agreement with the average time of the interactions found in our simulated sample. \\

We therefore propose backsplash orbits as a new mechanism to explain the presence of quiescent UDGs in low-density environments. This population of red and diffuse dwarfs results from the infall of normal star-forming field UDGs into galactic, group or cluster-size haloes responsible for stripping off their gas and propelling them to distances $\sim 1$ Mpc and beyond. In the most extreme cases, UDGs may even be ejected several Mpc away from these systems. The predicted fraction of red UDGs in our simulation in the studied mass range is $ \sim 24 \% $. Mild clustering and old stellar populations, along with dark matter haloes of lower mass and the complete absence of gas in the galactic and circumgalactic region, are the expected telltales of this formation scenario for red UDGs. Future wide-field surveys targeting the surrounding of groups and clusters may be the most promising way to uncover this population of elusive dwarfs predicted as a natural consequence of the assembly of haloes in $\Lambda$CDM.


\section*{Methods}
\label{sec:met}

\textbf{Simulation data and property determination}. For our calculations, we use the cosmological hydrodynamical TNG50 simulation \cite{Pillepich2019, Nelson2019TNG50} of the IllustrisTNG project \cite{Pillepich2018a, Pillepich2018b, Springel2018, Nelson2018, Naiman2018, Marinacci2018, Nelson2019TNG}. The IllustrisTNG galaxy formation model presents improvements in physics due to the effects of active galactic nuclei and feedback \cite{Weinberger2017, Pillepich2018a} compared to the original Illustris, its predecessor \cite{Vogelsberger2014a, Vogelsberger2014b, Genel2014}. The simulation was run using the moving-mesh code {\sc arepo} \cite{Springel2010,Weinberger2020}. Besides gravity, the runs model a set of physical processes relevant to galaxy formation including gas cooling and heating, star formation, metal enrichment, stellar and black hole feedback and magnetic fields. \\

The simulation is initialized at redshift $z=127$ using the N-GENIC code \cite{SpringelNGENIC} with cosmological parameters consistent with the Planck mission results \cite{PlankColaboration2016}: $\rm{ \Omega_m = \Omega_{DM} + \Omega_b = 0.3089 }$, $\rm{ \Omega_b = 0.0486}$, cosmological constant $\rm{ \Omega_{\Lambda} = 0.6911 }$ , Hubble constant $\rm{ H_0 = 100 \, h \, km \, s^{-1} \, Mpc^{-1} }$, with $\rm{h=0.6774}$, $\rm{ \sigma_8 = 0.8159 }$ and spectral index $\rm{ n_s = 0.9667 }$. \\

TNG50 is the smallest box from the TNG suite (50.7 Mpc (comoving) on a side compared to 100 Mpc and 300 Mpc) but the one with the highest resolution, a total of  $ 2160^3 $ gas and dark matter particles are set in the initial conditions, resulting in a mass per particle $\rm{ m_{bar} } = 8.4 \times 10^4 \, \rm {M_{\odot}} $ and $\rm{ m_{DM} }= 4.6 \times 10^5 \, \rm {M_{\odot}}$ for the baryons and dark matter, respectively. The gravitational softening for the stars and dark matter is $0.29$ kpc (comoving), whereas the gas has adaptive softening down to $74$ pc (physical). TNG50 is the only simulation of its kind and resolution that is able to follow such a wide range of environments from dwarf haloes to clusters of galaxies. The TNG50 box includes one (1) halo with $\rm{ \log (M_{200} / M_{\odot})> 14 }$ and a substantial number of less massive haloes ($\rm{ 13 < \log (M_{200} / M_{\odot}) < 14 }$) that allows the analysis of galaxies in the high-density environments of groups and clusters.\\

The identification of groups is done via the friends-of-friends (FoF) \cite{Davis1985} algorithm followed by {\sc subfind} to identify substructure \cite{Springel2001}. Subhaloes containing a stellar component are considered galaxies. Galaxies are classified either as `centrals' (main galaxy in each group) or `satellites' otherwise. Here we use centrals to identify the population of dwarf galaxies in the field, meaning they are not satellites of any more massive system. Galaxy quantities such as stellar mass $\rm{M_\star}$, morphology, age and star formation rate are defined using particles within the `galactic radius', which is defined as twice the half-mass radius of the stars: $\rm{r_{gal} = 2 \, r_{h\star}}$. Luminosities (used to compute colours) correspond to all stellar particles assigned to the galaxy (field SubhaloStellarPhotometrics in the halo catalog, \cite{Nelson2015}).\\ 

The morphology parameter $\kappa_{\rm rot}$ is calculated following [\cite{Sales2012}] as follows. After rotating each galaxy to a reference frame where the angular momentum of the stars (within $r_{\rm gal}$) points along the $z$ direction, we compare the energy in rotation around the $z$ axis to the total kinetic energy $K$ as $\kappa_{\rm rot} = (1/K) \Sigma\;  (1/2 m j_z^2/R)$, where $j_z$ is the angular momentum of each stellar particle in the rotated system, $m$ is their mass, $R$ is their cylindrical radii and the sum is over stars within $r_{\rm gal}$. Defined in this way, the morphology parameter $\kappa_{\rm rot}$ has been shown to correlate with other definitions of galaxy morphology \cite{Sales2012,Snyder2015}. Low $\kappa_{\rm rot}$ values correspond to spheroid-dominated objects, whereas $\kappa_{\rm rot}>0.6$ is used to identify disk-dominated objects. \\

Galaxies are followed over time by means of Sublink merger trees \cite{RodriguezGomez2015}. This allows us to track the mass, size and star formation histories of our sample over time. Note that the circumgalactic gas properties in galactic-size and group-size haloes in TNG50 are in good agreement with observational constraints \cite{DeFelippis2021, Nelson2019TNG50} and may therefore provide a solid theoretical ground to study environmental effects in our UDG sample.\\

The stellar mass for the lenticular galaxy NGC 1211 was estimated from its V-band luminosity in [\cite{Fuse2012}] and assuming a mass to light ratio of $1$ for simplicity.\\
\newline
\textbf{Sample of field UDGs}. The criterion to define UDGs varies across different works in the literature. Here we define UDGs as the most extended outliers of the stellar mass-size relation, following the philosophy introduced in [\cite{Lim2020}]. The galaxy modeling used in the TNG100 and TNG50  simulation has been shown to agree well with observational constraints on the stellar mass-size relation of the galaxy population \cite{Genel2018,Pillepich2019}. In this work, we construct the mass-size relation using well resolved galaxies, defined as those with dark matter mass $\rm{ m_{DM} \geq 5 \times 10^7 \, M_{\odot} }$ (with total dark matter mass assigned by {\sc subfind} to each subhalo), stellar mass $\rm{ m_{\star} \geq 5 \times 10^6 \, \rm M_{\odot} }$ and size $\rm{ r_{h\star} \geq 0.3 \, \rm kpc }$; this results in a minimum number of $\sim 60$ stellar and $110$ dark matter particles.\\ 

Fig.~\ref{fig:sample} in the main text shows the stellar mass-size relation, where the median at fixed $\rm{M_\star}$ is indicated by the thick black line. We notice that for $\rm log(M_\star/M_\odot) < 7.5$ the median starts to steadily increase towards lower-mass objects, an effect not found in observation and of origin likely numerical. To be conservative, we study only dwarf galaxies in the stellar mass range $\rm{ \log (\rm{M_{\star} / M_{\odot}) = [7.5, 9] }}$. More than $\rm{8600}$ dwarf galaxies in TNG50 satisfy this mass cut. The distribution of sizes at a given stellar mass is approximately log-normal. We, therefore, select the $5\%$ most extended outliers at fixed $M_\star$ as our UDG population, deeming all galaxies within $5$th- $95$th percentiles as `normal' galaxies. This results in an average half mass radius $\rm{2.5 \pm 0.8 \, kpc}$ for our UDG population. From the UDGs identified in this way, $176$ are centrals to their haloes (or `field' population) and constitute the sample analyzed here (the study of UDGs as satellites is presented elsewhere by J.A.B. et al., manuscript in preparation). Fig.~\ref{fig:sample} shows that our definition for UDG galaxies is in very good agreement with several observational samples of star-forming and quenched UDGs in low density environments \cite{Rong2020b, MartinNavarro2019, Roman2019}.\\
\newline
\textbf{Visualizations}. Images shown in Fig.~\ref{fig:orbit_box} were made using the Py-SPHViewer code\cite{Benites2015} v1.0.0. This code smooths the particle information into two-dimensional histograms to reflect the underlying continuous density field. For the specific case of the small panels in the small panels in Fig.~\ref{fig:orbit_box}, we combined the information from the gas cells (blue hues) and the stellar particles (red). Each stamp has $150$x$150$ pixels and we use $12$ neighbors for the smoothing. We use all gas or stellar particles identified to belong to this subhalo by {\sc subfind} which are within the image box ($12$ kpc (physical) on a side). For the bottom panel we use the XY coordinates of each subhalo in the halo catalog. The image is smoothed using a $3$ neighbor kernel density estimation and has $1000$x$500$ pixels.\\ 


\section*{Acknowledgements}
The authors would like to thank the referees for useful and constructive reports that helped improve this manuscript.
JAB and MGA acknowledge financial support from CONICET through PIP 11220170100527CO grant. LVS is grateful for support from the NSF-CAREER-1945310 and NASA ATP-80NSSC20K0566 grants. 
AP acknowledge support from the Deutsche Forschungsgemeinschaft (DFG, German Research Foundation) – Project-ID 138713538 – SFB 881 (“The Milky Way System”, subproject C09). DN acknowledges funding from the Deutsche Forschungsgemeinschaft (DFG) through an Emmy Noether Research Group (grant number NE 2441/1-1). FFM acknowledges support through the program "Rita Levi Montalcini" of the Italian MUR. 
MC is partially supported by NSF grants AST-1518257 and AST-1815475. PT acknowledges support from NSF grants AST-1909933, AST-200849 and NASA ATP grant 80NSSC20K0502. MV acknowledges support through NASA ATP grants 16-ATP16-0167, 19-ATP19-0019, 19-ATP19-0020, 19-ATP19-0167, and NSF grants AST-1814053, AST-1814259,  AST-1909831 and AST-2007355.

\section*{Author contributions}
The listed authors have made substancial contributions to this manuscript, all coauthors read and commented on the document. JAB led the analysis of the simulation, compiled observational data from the literature and made all figures. LVS and MGA are responsible for the original idea and mentorship of JAB throughout the project. LVS led the writing of the manuscript and the response to the referee report with substantial contributions from JAB, MGA, AP, DN, FM and LH. Coauthor MC provided the expertise on the observational consequences of the results and on the study of quenching of dwarf galaxies. AP, DN, FM, RP, PT, MV and LH are core members of the TNG50 simulation who set-up, developed and run the simulation this manuscript is based on. 

\section*{Data availability}
This letter is based on snapshots, subhalo catalogs and merger trees from the cosmological hydrodynamical TNG50 simulation \cite{Pillepich2019, Nelson2019TNG50} of the IllustrisTNG project \cite{Pillepich2018a, Pillepich2018b, Springel2018, Nelson2018, Naiman2018, Marinacci2018, Nelson2019TNG}. These data is publically available at \href{https://www.tng-project.org/}{https://www.tng-project.org/}. Ascii tables with the simulation data for our sample of UDGs in Figs.~\ref{fig:sample},~\ref{fig:color} and ~\ref{fig:features} are available in the public repository: \href{https://github.com/josegit88/public_data_files/tree/main/ascii_files_isolated_UDGs_TNG50}{here}. Source Data are provided with this paper. 

\section*{Code Availability}
Scripts used for reading/access to the snapshot, merger trees and subhalo data are publically available at the \href{https://www.tng-project.org/data/docs/scripts}{TNG database}. Visualizations were made using the publically available Py-SPHViewer code \cite{Benites2015}. Any correspondence and/or request for materials pertaining to this manuscript should be directed to J.A.B.

\section*{Competing interests}
The authors declare no competing interests.

\section*{Additional information}
Supplementary information is available for this work.


\section*{Supplementary Notes}
We collect here supplementary information to support our analysis presented in the Letter and Methods sections of this manuscript. The supplementary materials include 6 additional figures with their related discussion and references. \\

\section*{Supplementary Information}

Backsplash orbits may place galaxies well beyond the virial radius of their host halo, in some cases far enough that the subhalo finder (in this case {\sc subfind}) identifies them as isolated/central galaxies again. Satellites in such orbits are needed to reproduce the observed fraction of quiescent galaxies as a function of radius in galaxy groups and clusters out to at least twice the virial radii \cite{Balogh2000, Mamon2004, Gill2005, Wetzel2014}. This backsplash mechanism has been found to be more efficient for low mass subhaloes \cite{Sales2007b, Ludlow2009}, therefore in the dwarf galaxy population. For instance, quiescent dwarfs outside the virial radius of the Milky Way or Andromeda in the Local Group may be explained as backsplash objects \cite{Fillingham2018}. 

\begin{figure*}[!ht]
\centerline{
\includegraphics[width=\linewidth]{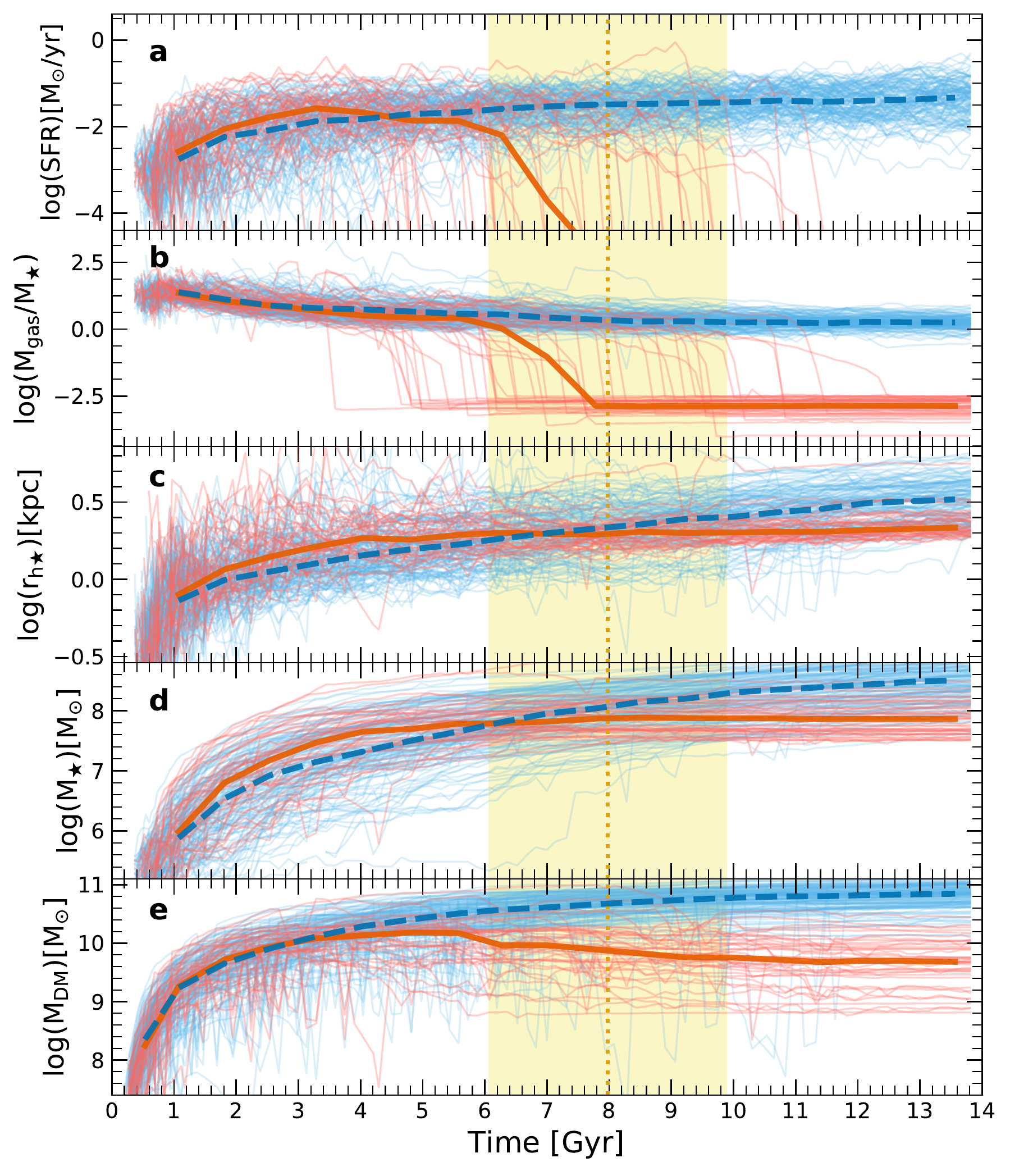}
}
\caption{{\bf Time evolution of galaxy properties for quiescent and star-forming UDGs}. From top to bottom: star formation rate ($\rm{SFR}$, \textbf{a}), gas fraction defined as $\rm{M_{\rm gas}/M_{\star}}$ ratio (\textbf{b}), stellar size ($\rm{r_{h \star}}$, \textbf{c}), stellar mass ($\rm{M_{\star}}$, \textbf{d}) and dark matter mass ($\rm{M_{DM}}$, \textbf{e}). Thin lines indicate individual galaxies while the thick continuous and dashed curves (red and blue respectively) correspond to medians of each UDG sample at a given time. The orange dotted vertical line correspond to the average time of infall for quiescent UDGs, while the shadowed yellow region indicates its standard deviation (see Fig.~\ref{fig:hist_times}).}
\label{fig:evol_features}
\end{figure*}

Previous works linking the quenching of galaxies to backsplash orbits around groups and clusters have been analytical, semi-analytical or based on N-body only simulations. Only recently, by using the cosmological hydrodynamical simulation TNG50 it has been possible to follow self-consistently the physical processes leading ultimately to the gas removal and quenching of backsplash objects, for which high resolution is key in order to resolve not only the internal structure of galaxies, but also the structure of the circumgalactic medium causing the gas stripping \cite{Nelson2020}. The role of backsplashorbits on the quenching of the dwarf population as a whole was studied in TNG50\cite{Joshi2021}, finding that by excluding backsplash galaxies the fraction of quenched dwarfs in the field with $M_\star \geq 10^8 \; \rm M_\odot$ is practically zero, in good agreement with observational estimates from SDSS \cite{Geha2012} and at most $\sim 5\%$ at the lowest stellar mass considered here, $M_\star=10^{7.5}\; \rm M_\odot$.

\begin{figure*}[!ht]
\centerline{
\includegraphics[width=0.9\textwidth, trim=0.25cm 0.25cm 0.25cm 0.2cm,clip=true]{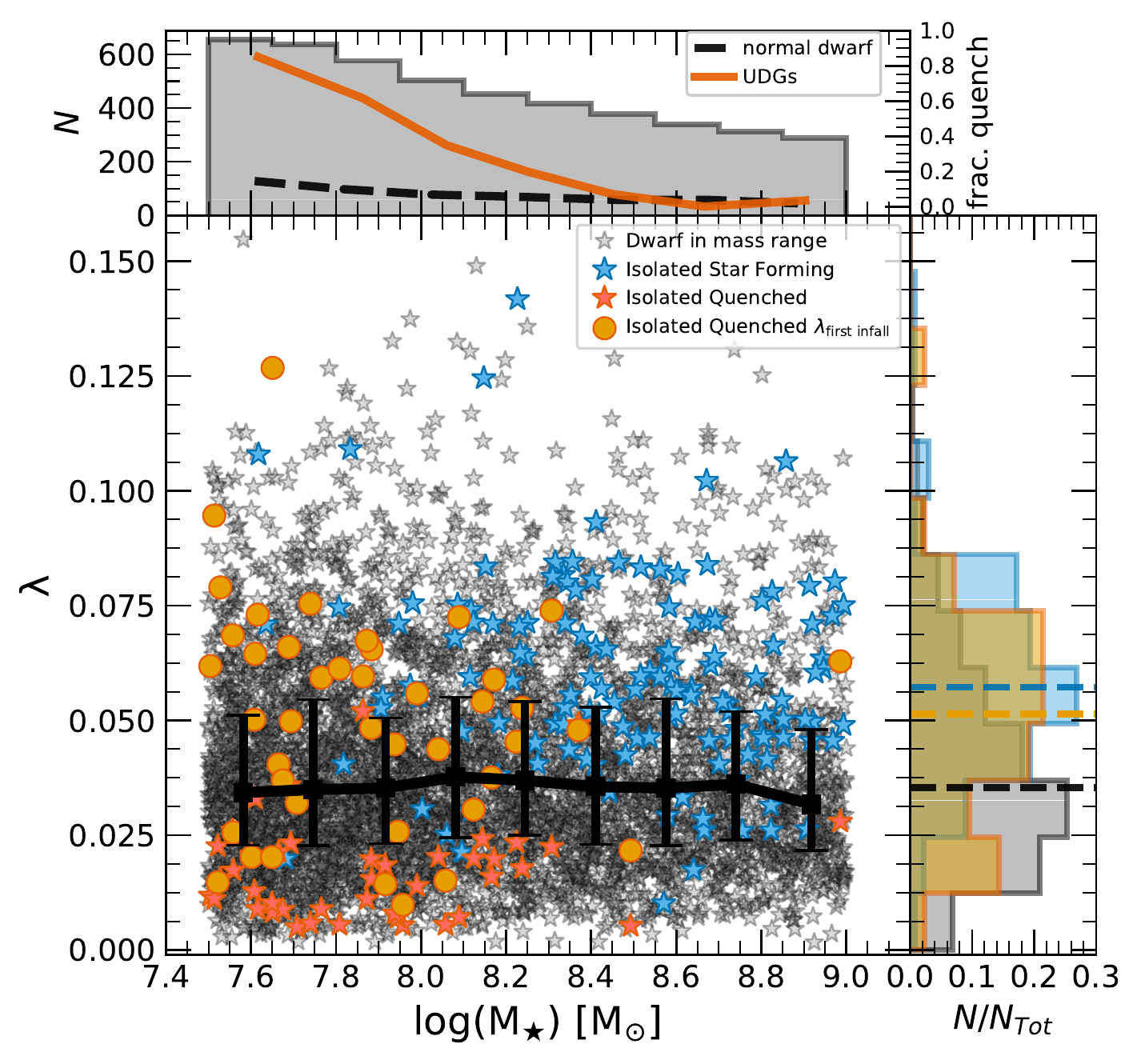}
}
\caption{{\bf Formation mechanism of UDGs in the field}. Halo spin for dwarf galaxies at a given stellar mass. Normal field dwarfs are shown in gray, while stars correspond to field star-forming UDGs (blue) and quiescent UDGs (red) measured at $\rm{z=0}$ and at infall (orange circles). The median spin at fixed $\rm{M_\star}$ of the normal dwarf population is indicated by the solid black curve and with error bars indicating 25th-75th percentiles, the average value for all mass bins is $\lambda_{\rm{dwarf}} = 0.035^{+0.017}_{-0.012}$. UDGs occupy preferentially higher-spin haloes. The median and rms dispersion for the UDGs are $\lambda_{\rm{blue}} = 0.059 \pm 0.021$ and $\lambda_{\rm{red, \, infall}} = 0.051 \pm 0.024$, for the blue and red population, respectively. Red UDGs are shown at infall (orange circles) and for comparison at $\rm{z=0}$ after the interaction with their hosts, which lowers their angular momentum via tidal stripping $\lambda_{\rm{red,z=0}} = 0.017 \pm 0.011$.  The different halo spins in normal vs. UDG population is more clearly seen in the histograms on the right, with dashed lines indicating the medians of the normal, blue UDGs and red UDGs (at infall). The top panel shows, in addition to the stellar mass distribution of the dwarf population (gray histogram), the quenched fraction of field UDGs (red) and of field normal dwarfs (black line).}
\label{fig:spin}
\end{figure*}

Interestingly, the fraction of quiescent field UDGs in our sample is larger than taking the whole dwarf population: we find that 25\% of field UDGs in the field are quenched when averaged in our whole mass range, but up to 50\% in the lowest mass half. For comparison, the quenched fraction for the normal dwarf population in the field is $7.8\%$ in the whole mass range and $10.2\%$ in the lowest mass half (see upper panel in Fig.~\ref{fig:spin}). Restricting the analysis to only backsplash objects, $100\%$ of backsplash UDGs are quenched while this fraction increases from $70\%$ (for the low mass end) to $100	\%$ (high mass end explored here) in normal dwarfs. Although it is tempting to interpret these numbers as UDGs being more susceptible to quenching during their backsplash trajectories compared to normal dwarfs, we find that, instead, the difference is due to a different distribution of host halo masses with which normal and UDGs galaxies have interacted in the past.\\ 

\begin{figure*}[!ht]
\includegraphics[width=\linewidth]{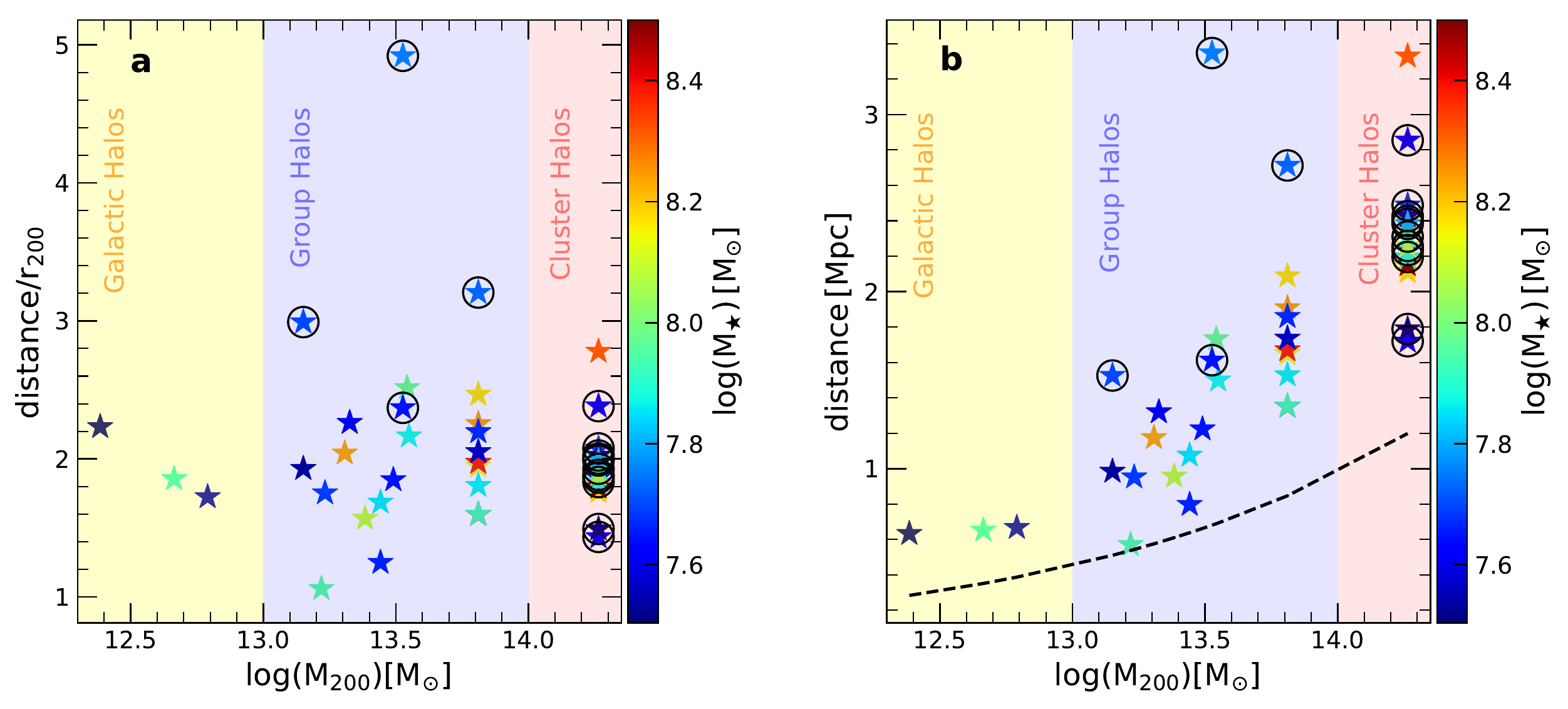}\par 
\caption{{\bf Location of backsplash UDGs in the field as a function of their past host halo virial mass}. In panel \textbf{a}, present-day distance is shown in units of the virial radius of the past host halo, with a mean of $\rm{ \sim 2.1 \pm 0.6 \times r_{200}}$ for the sample. Panel \textbf{b}: same as before, but now distance is shown in Mpc, the dashed black line indicates the average virial radius ($\rm{r_{200}}$) at a given virial mass of the host. Encircled in black are the UDGs that were ``pre-processed" and quenched in haloes with an average $\rm{M_{200} \sim 9.16 \times 10^{12} M_{\odot}}$ before falling into the last host halo they interacted with and that placed them on backsplash orbits. The colour code refers to stellar masses of each quenched UDGs at $\rm{z = 0}$ and the shaded regions limit the type of halo they are backsplash galaxies of (galactic, groups or cluster haloes).}
\label{fig:distances}
\end{figure*}

For instance, for backsplash objects that interacted with $M_{200} \geq 10^{13}\; \rm M_\odot$ hosts, the fraction of quenched normal dwarfs and UDGs is, in both cases, $ > 90\%$. Normal dwarfs that interacted with lower-mass hosts ($M_{200} < 10^{13}\; \rm M_\odot$) are the ones showing a $50\%$ quenched fraction and driving down the overall quiescent fraction for the normal population. However, we have only $3$ UDGs interacting with galactic halos and therefore conclusive claims on whether the extended structure of UDGs may turn them more susceptible to ram-pressure stripping and other environmental effects will have to wait until larger volume simulations of this kind provide a more solid statistical basis for such a study.\\

\begin{figure}
	\includegraphics[width=\columnwidth]{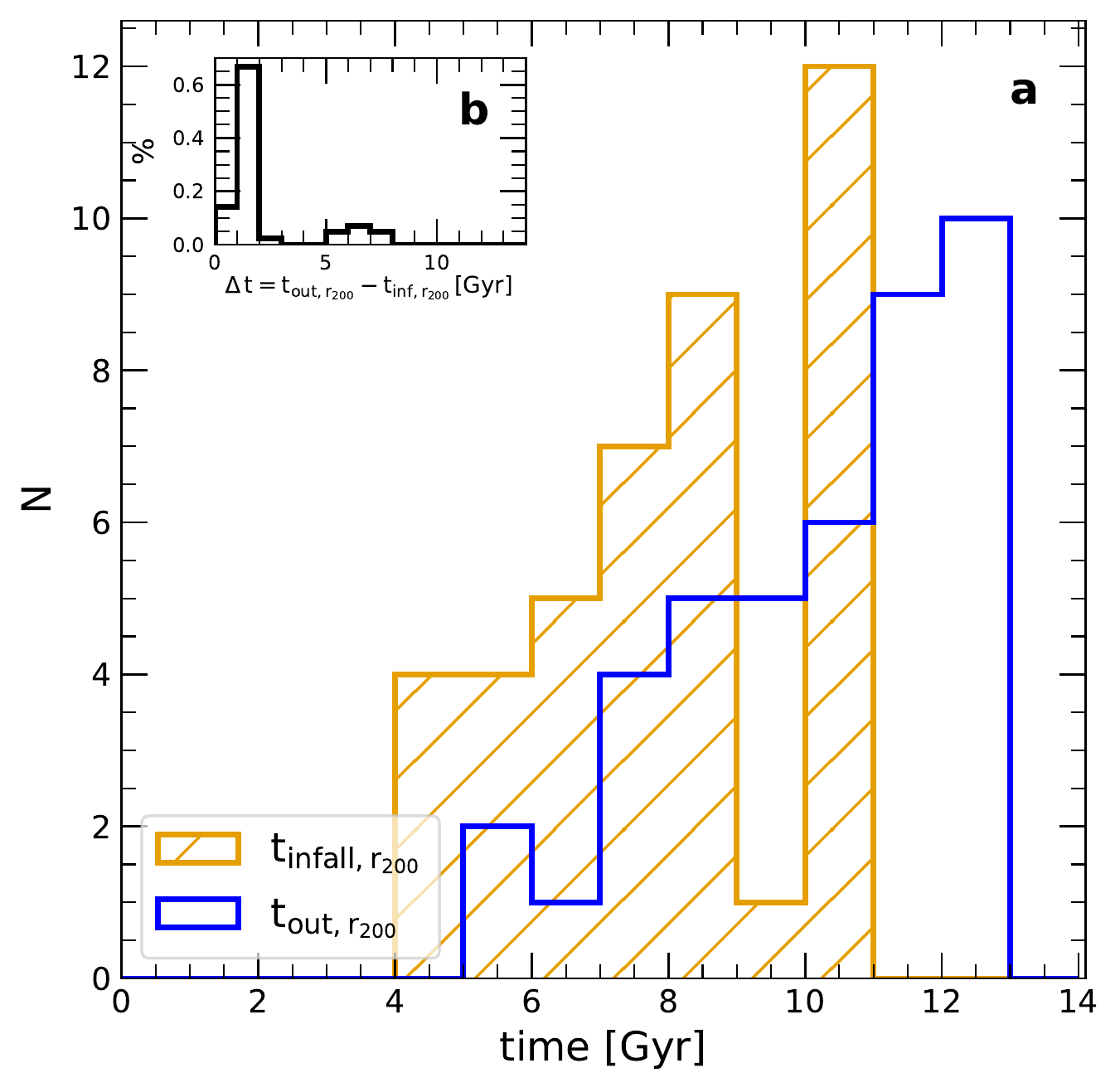}
    \caption{{\bf Distribution of the infall and output times of field red UDGs}. Panel \textbf{a}: infall (orange) and output (blue) times are defined as the time each object crosses the virial radius (inwards or outwards respectively) of the host that placed them on backsplash orbits. Panel \textbf{b}: distribution of time interval between infall and exit from the host. On average these UDGs fell $\rm{5.83 \pm 1.92}$ Gyr ago and stayed only $\rm{\sim 1.5}$ Gyr (median) within their temporary host haloes.}
\label{fig:hist_times}
\end{figure}

This highlights that, when studying the tail of dwarfs with the most extended radii, the importance of backsplash orbits and of its associated quenching is enhanced with respect to the normal population of dwarfs. We emphasize that in the case of our quiescent UDGs, backsplash orbits are responsible for quenching and placing UDGs back in the field, but they play a non-dominant role on setting their extended sizes: quenched UDGs in the field were already diffuse and extended before infalling into their past hosts.\\

We present this in more detail in Fig.~\ref{fig:evol_features}, where we show several galaxy properties as a function of time for our population of field UDGs, star-forming (blue) and those quiescent (red). Individual galaxies are shown with thin lines, while medians are highlighted in thick curves. The yellow shaded region indicates the average infall time of the red UDG population. The middle panel shows that red UDGs sizes were already extended before their average infall time and that the interaction had little impact afterwards. On average, our quenched UDG sample increased their size only by 25\% compared to their infall stellar half-mass radius. The top and second rows nicely illustrate that at the time of interaction with the hosts is when the removal of gas occurs followed by their quenching. This is in good agreement with the galaxy transformations experienced by satellites in the TNG simulation suite \cite{Yun2019,Stevens2019,Stevens2021}. The quenching experienced in red UDGs around $t \sim 6$ Gyr means that their early stellar growth should be systematically faster than their star-forming counterparts (see d panel in Fig.~\ref{fig:evol_features}), which continue to form stars until present day \cite{Mistani2016}.\\

This still begs the question: what made quiescent UDGs be so extended in the first place? We find that in TNG50, the UDG population (red and blue) forms in dwarf haloes with biased-high spins, and it is this excess of angular momentum compared to the normal population what is responsible for their extended sizes (see Fig.~\ref{fig:spin}, blue and orange symbols). Spins are calculated  using the definition introduced in [\cite{Bullock2001}]. Note that quiescent UDGs should be considered at infall to study the origin of their sizes. At present-day, their interactions with a past host has stripped off the outer halo layers lowering their initial angular momentum content (see red stars). A Kolmogorov-Smirnov test to compare the spin distribution of normal (gray points, black curve) vs. UDGs (orange/blue symbols) retrieves a p-value $=2.63 \times 10^{-7}$, confirming that $\lambda$ values in both samples are not drawn from the same parent distribution. Large spins as origin for the extended sizes in UDGs is in excellent agreement with one of the first theoretical models of UDG formation presented \cite{Amorisco2016} and with recent kinematical modeling of observed UDGs in the field \cite{Mancera-Pina2020}. \\

\begin{figure*}
\centering
\includegraphics[width=0.75\textwidth]{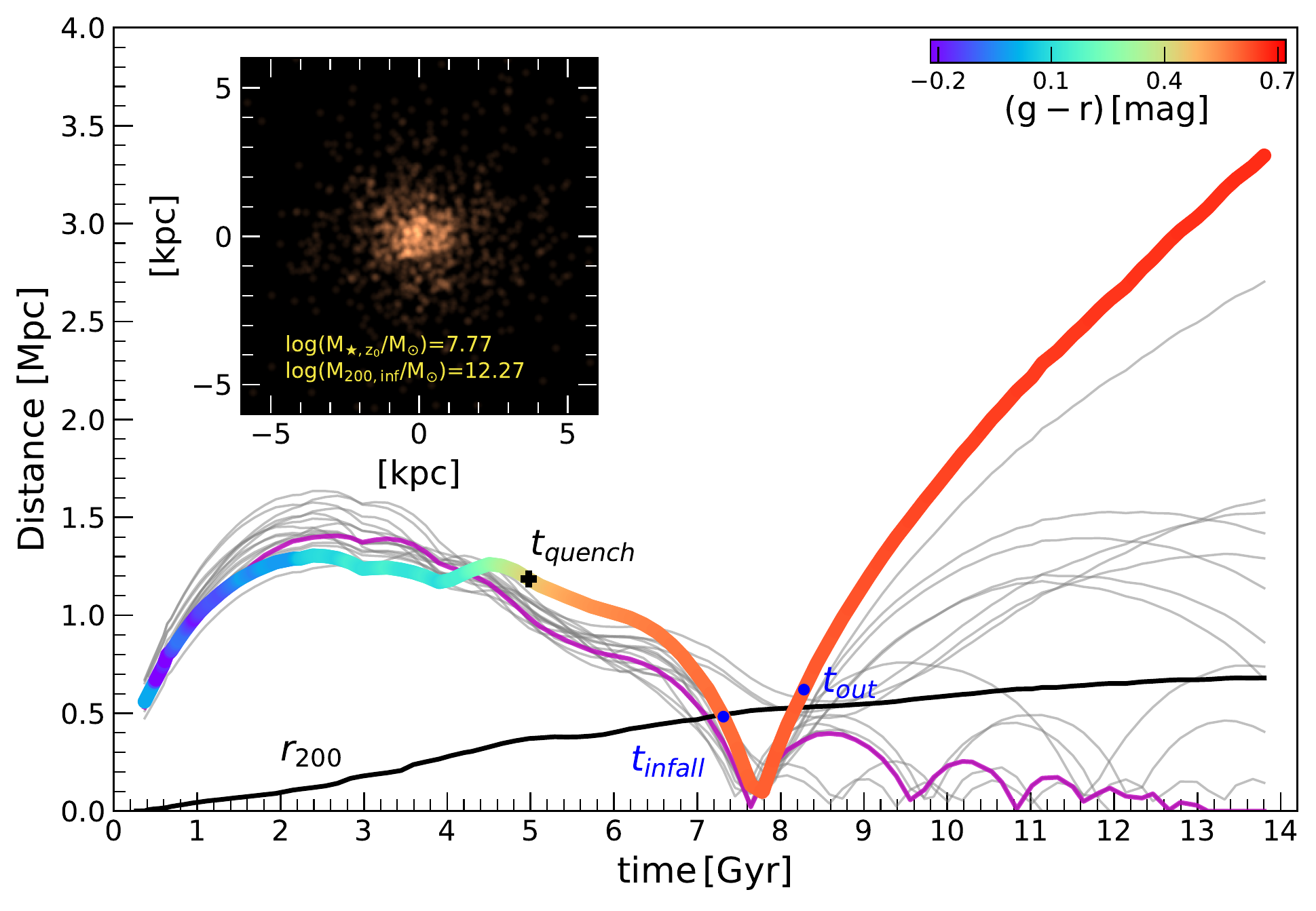}
\caption{{\bf Example of our most extreme object: a quiescent UDG at $3.35$ Mpc ejected from its host}. The thick coloured line shows the distance of this specific UDG from its last group-sized host from which it is ejected at $t \sim 7.5$ Gyr. As shown in Fig.~\ref{fig:orbit_box} of the main text, the orbit is colour coded according to the instantaneous ($g$ - $r$) galaxy colour (see colour bar). The black curve indicates the virial radius of the host group. The UDG infalls into this group as part of an association of satellites in a galaxy-size group with $\rm{M_{200} \sim 10^{12} M_{\odot}}$ which is responsible for its quenching (``pre-processing") $\sim 9$ Gyr ago. Gray lines show the orbits of the other 17 satellites in the galactic group and magenta line shows the central of this group. The small snapshot view shows the UDG at present-day.}
\label{fig:ejected}
\end{figure*}

Interestingly, contrary to this theoretical prediction, other simulations have found little dependence of their simulated field UDGs on halo spin, citing instead burstiness \cite{DiCintio2017, Chan2018} and mergers \cite{Wright2021} as the origin for their extended sizes. For the high density environments of groups and clusters, additional mechanisms have been proposed to alter the size and the gas content of cluster UDGs \cite{Safarzadeh2017,Carleton2019,Jiang2019,Sales2020}. We present a detailed study of the formation of UDGs in TNG50 and their relation to environment in a companion paper (Benavides et al., in-prep). Here, we focus instead on possible mechanisms to explain a specific sub-population of UDGs: those that are quiescent and in the field today.\\

In Fig.~\ref{fig:distances} we show for each quiescent UDG the present-day distance from the halo they backsplash from as a function of the present day virial mass of such haloes. The left panel shows that, on average, quenched UDGs are found at $r \sim 2.1 r_{200}$ (in good agreement with N-body only predictions \cite{Mamon2004}), but may reach as much as $4.9 r_{200}$  ($3.35$ Mpc). In the right panel, distances are shown in kpc. Dwarfs interacting with galactic haloes ($\rm{M_{200}<10^{13} \; \rm M_\odot}$) may be found today $\sim 650$ kpc away from them. For groups and clusters, distances $\ge 1$ Mpc are common. Moreover, because of their foreign origin, the relative velocities of backsplash UDGs with respect to their surrounding can be significant. On average, quenched UDGs within 1 Mpc of their host today have a relative velocity $\Delta v \sim 200$ km/s with respect to any other surrounding galaxy (here we define the center of velocity by using all galaxies with $M_\star \ge 10^{10}\; \rm M_\odot$ within $1$ Mpc). The relative velocity though increases with distance to their past hosts, reaching on average $\Delta v \sim 500$ km/s (and individually up to $\sim 1000$ km/s) if the red UDG is located today more than 1 Mpc away from its host. This is interesting in light of the observed quenched UDG S82-DG-1, which shows a high peculiar velocity\cite{Roman2019}.

For most quiescent UDGs, the last system they interacted with -- and that placed them in backsplash orbits-- is the same as that  responsible for their star formation shutting off. However, a non-negligible fraction ($\rm{ \sim36\% }$), are quenched as result of ``pre-processing", mechanism quantified in the TNG simulations\cite{Donnari2021}. We highlight such UDGs with a black circle in Fig.~\ref{fig:distances}. The median virial mass of the halo where quenching in these galaxies happens is $\rm{M_{200} = 9.16 \times 10^{12} \; \rm M_\odot}$ (measured at the time of quenching of each dwarf). This confirms that galaxy-mass haloes may also be capable of forming and hosting quiescent UDGs, extending towards lower densities, beyond groups and clusters, the range of environments where red UDGs are found in large numbers.

In order to reach large distances from their hosts, the interactions placing UDGs on backsplash orbits must have occurred some time ago. In Fig.~\ref{fig:hist_times} we show the distribution of infall times for all red UDGs (orange), where infall is defined as the first time they crossed the virial radius of the FoF group they interacted with and is responsible for their placement on external orbits. The distribution is wide, but on average red UDGs interacted with their hosts at $\rm{\sim 5.83 \pm 1.92}$ Gyr ago (lookback time) or redshift $\rm{z \sim 0.6}$. The interaction in most cases is brief and corresponds to only a single pericentric passage. The blue histogram in Fig.~\ref{fig:hist_times} shows the distribution of departure times from the interaction group for each red UDG, $t_{\rm out}$, defined as the last time when they cross (outwards) the $\rm{r_{200}}$ of the host. The small inset panel shows that most UDGs spend less than $2$ Gyr within their hosts (median $1.5$ Gyr). \\

The most extreme object in our sample is found today beyond $3.35$ Mpc ($\rm{\sim 4.9 \times r_{200}}$) from the group it interacted with $\rm{\sim 6}$ Gyr ago. Its unusual orbit, one in which the apocenter is larger than the turnaround radius, results from its common infall as part of a galaxy-sized group ($\rm{M_{200} = 1.86 \times 10^{12} \; M_\odot}$ at $t=6$ Gyr) hosting multiple satellites itself (17 with $\rm{M_\star > 5 \times 10^6 \; \rm M_\odot}$). Fig.~\ref{fig:ejected} shows the orbit of this specific quiescent UDG (thick coloured line) along with those of all galaxies in the galaxy-mass halo (gray lines for satellites, magenta for the central) that fall together into the final group host with $\rm{M_{200}(z=0) = 3.36 \times 10^{13} \; \rm M_\odot}$. A snapshot view of the red UDG at the present day is included. \\

The UDG is first quenched in the galaxy-mass system (``pre-processing") to later fall into the group and gain energy from multiple-body interactions at the time of the first pericenter \cite{Sales2007b, Ludlow2009}. This results in an unbound orbit and its extreme isolation at the present day. Although most of the backsplash galaxies will not result in unbound orbits, some extreme free-floating cases like the one presented here are to be expected within $\Lambda$CDM. Moreover, these unorthodox orbits are increasingly more common among low mass haloes \cite{Ludlow2009}, suggesting that a population of low mass red UDGs may be lurking in low density regions of the universe awaiting to be discovered. Some observational evidence of UDGs clustered around groups and clusters already exists \cite{Roman2017a,Roman2017b}. \\

\begin{figure}
	\includegraphics[width=\columnwidth]{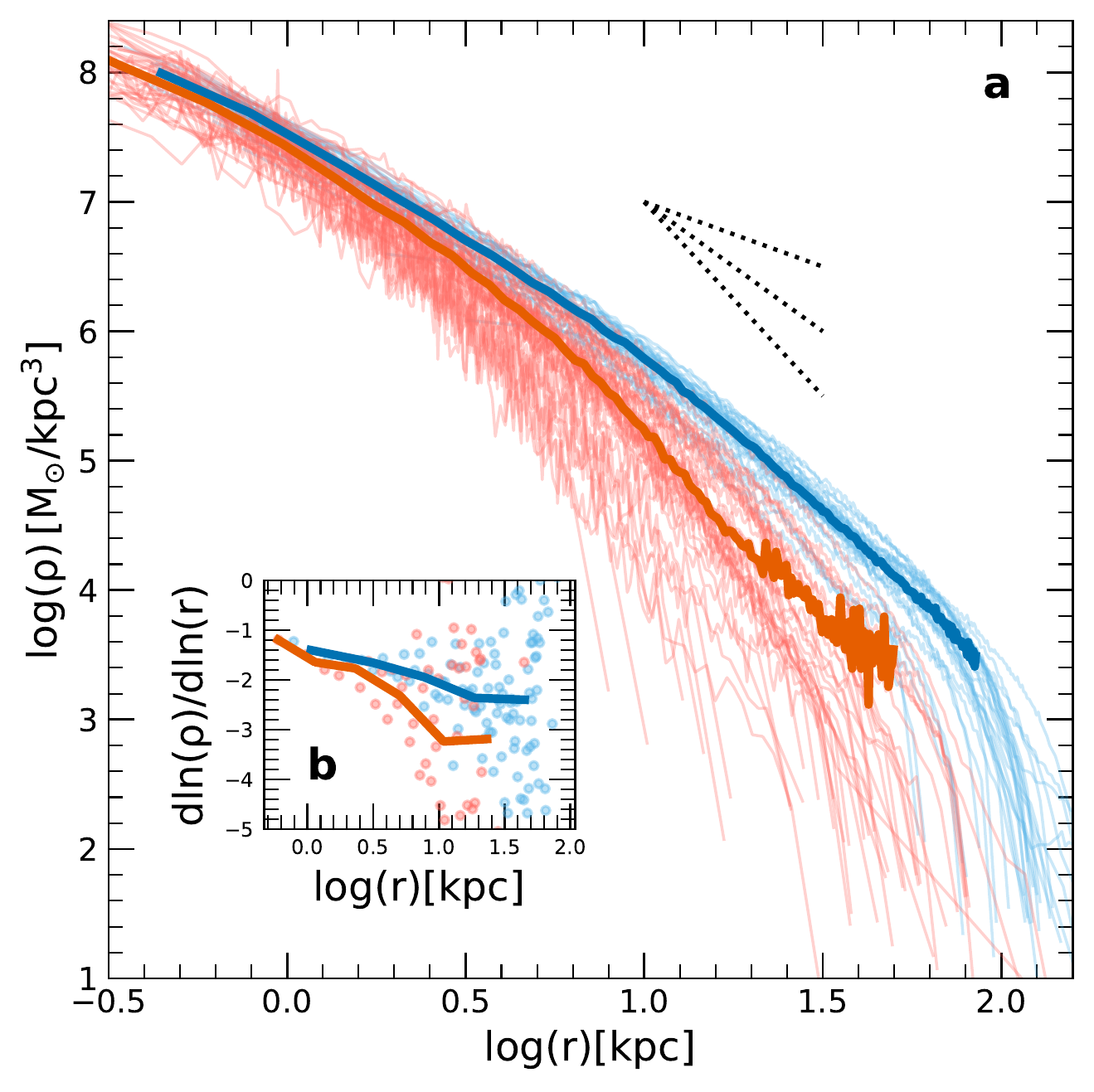}
    \caption{{\bf Dark matter density profile of red vs. blue UDGs}. Panel \textbf{a}: as a result of the interaction with their temporary host, red UDGs are predicted to have been tidally stripped showing today steeper outer density profiles compared to similar-mass blue UDGs. Thin lines show individual objects while medians are indicated by thick lines. The dotted lines serve as a guide for slopes: $\alpha = -1, -2, -3$. Panel \textbf{b}: the small inset shows the logarithmic slopes of the profiles in the main panel as a function of distance. Symbols correspond to individual objects and the median trend is shown by the thick curves. Dark matter profiles of red UDGs remain similar to field blue UDGs in the inner regions.}
\label{fig:rho_dm}
\end{figure}

Besides loosing all their gas to environmental effects during the interactions with their once hosts, red UDGs experience also tidal stripping of their outer dark matter haloes. This is shown in Fig.~\ref{fig:rho_dm}, where thin red/blue lines show the dark matter density profile of quenched and star-forming UDGs, respectively. Matching coloured thick curves highlight the median profile of each sample. Field red UDGs have on average dark matter haloes more steeply declining with radius than the blue population, resulting in smaller virial masses at the same stellar mass, as shown in the right panel of Fig.~\ref{fig:features} in the main text. This is in agreement with the effects of tidal interactions of satellite galaxies in TNG \cite{Engler2021}.  \\

The small inset in Fig.~\ref{fig:rho_dm} shows the slopes of the profiles (dots for individual curves, thick red/blue line for the median of quenched and star-forming UDGs). Differences between the quiescent and star-forming UDG population are significant only in the outer regions, where tidal stripping operates, while both populations are indistinguishable in the inner regions \cite{Zahid2018}. This means that no major differences may be expected in mass (or velocity) estimates of star-forming and quiescent UDGs based on centrally concentrated tracers, such as stars or inner globular clusters. \\


\begin{flushleft}
{\bf \large References}
\end{flushleft}

\end{document}